\documentclass[aps,prl,twocolumn,showpacs,showkeys,superscriptaddress,a4paper]{revtex4}

\usepackage{graphicx,amssymb,amsmath,color}
\usepackage{dcolumn}% Align table columns on decimal point
\usepackage{bm}

\begin{document}
\bibliographystyle{My}

\title{Fission studies with 140~MeV $\bm{\alpha}$-Particles}
%%%%%%%%%%%%%%%%%%%%%%%%%%%%%%%%%%%%%%%%%%%%%%%%%%%%%%%%%%%%%%%%%%%%%%%%%%%%%%%%%%%%%%%%%%%

\author{A.~Buttkewitz}
\affiliation{I.\ Institut f\"{u}r Experimentalphysik, Universit\"{a}t
Hamburg, Hamburg, Germany}

\author{H.~H.~Duhm}
\affiliation{I.\ Institut f\"{u}r Experimentalphysik, Universit\"{a}t
Hamburg, Hamburg, Germany}

\author{F.~Goldenbaum}
\affiliation{Institut f\"{u}r
Kernphysik, Forschungszentrum J\"{u}lich, J\"{u}lich, Germany}

\author{H.~Machner}
\email{h.machner@fz-juelich.de} \affiliation{Institut f\"{u}r
Kernphysik, Forschungszentrum J\"{u}lich, J\"{u}lich, Germany}
\affiliation{Fachbereich Physik, Universit\"{a}t Duisburg-Essen,
Duisburg, Germany}

\author{W.~Strau{\ss}}
\affiliation{I.\ Institut f\"{u}r Experimentalphysik, Universit\"{a}t
Hamburg, Hamburg, Germany}

\date{\today}

%%%%%%%%%%%%%%%%%%%%%%%%%%%%%%%%%%%%%%%%%%%%%%%%%%%%%%%%%%%%%%%%%%%%%
\begin{abstract}%
Binary fission induced by 140~MeV $\alpha$-particles has been
measured for $^{\rm nat}$Ag, $^{139}$La, $^{165}$Ho and $^{197}$Au
targets. The measured quantities are the total kinetic energies,
fragment masses, and fission cross sections. The results are compared
with other data and systematics. A minimum of the fission probability in the vicinity $Z^2/A=24$ is observed.
\end{abstract}

%%%%%%%%%%%%%%%%%%%%%%%%%%%%%%%%%%%%%%%%%%%%%%%%%%%%%%%%%%%%%%%%%%%%%
%%
%%%%%%%%%%%%%%%%%%%%%%%%%%%%%%%%%%%%%%%%%%%%%%%%%%%%%%%%%%%%%%%%%%%%%
%
\maketitle

%%\section{Introduction}\label{sec:Introduction}

The binary fission process in heavy elements has been systematically
studied with energetic probes such as photons, protons,
$\alpha$-particles, as well as with heavy ions. The details of such
experiments can be found in Refs.~\cite{Vanden73}
and~\cite{Wagemans91}. Less is known about the fission of lighter
nuclei and higher energies where such nuclei can fission. If the
angular momentum is high then the fission barrier is reduced and even
light systems like $^{60}$Zn can undergo fission~\cite{Oertzen08}.
Here we will concentrate on reactions induced by light charged
particle and in particular $\alpha$-particles. The low mass region is
interesting since it is predicted that for a fissility parameter
$Z^2/A$ below 20 the system tends to become asymmetric. This is the
so-called Businaro-Gallone point~\cite{Businaro57}. Nix and
Sassi~\cite{Nix66} found in calculations employing the liquid drop
model that the probability for fission had a minimum at the quoted
fissility parameter. Such a minimum corresponds of course to a
corresponding maximum in the height of the fission barrier.

In the present work we extend upwards the energy of
$\alpha$-particles in order to have higher partial waves involved in
the reaction. In this way one might expect to be more sensitive to
the predicted asymmetric instability.

%%\section{Experiments}\label{sec:Experiments}

The experiments were performed at the J\"{u}lich cyclotron. A beam of
$\alpha$-particles was focussed onto the fissile targets in the
center of a scattering chamber which was 1~m in diameter. The beam
was then dumped into a well shielded Faraday cup. The targets were
$^\text{nat}$Ag, $^{139}$La, $^{165}$Ho and $^{197}$Au with
thicknesses of 50~$\mu$g/cm$^2$, 120~$\mu$g/cm$^2$, 97~$\mu$g/cm$^2$
and 130~$\mu$g/cm$^2$, respectively. The rare earth targets had
backings of carbon with thicknesses of 10~$\mu$g/cm$^2$ (La) and
30~$\mu$g/cm$^2$ (Ho). The lanthanum target had in addition a
30~$\mu$g/cm$^2$ carbon coating to avoid oxidation.

Two different setups were used in the experiments. First, two solid
state detectors of 30~mm diameter, cooled to $-20^{\circ}$, were
mounted symmetrically left and right of the beam direction at angles
corresponding to full momentum transfer to the compound nucleus. The
solid angles were defined by collimators of 25~mm diameter. Whilst
the right detector was only 57~mm away from the target the left one
was at a distance of 150~mm. Thus, if a fragment was detected in the
smaller solid angle, its complementary fragment should always be
detected in the larger solid angle. Coincidence circuits ensured that
the two fragments were from the same reaction event. A high voltage
of 10~kV applied to the target holder prevented electrons from
reaching the detectors.

The detectors were calibrated with fission fragments from a
$^{252}$Cf source applying the method of Schmitt \emph{et
al.}~\cite{Schmitt65} and Kaufmann \emph{et al.}~\cite{Kaufmann74}.
The calibrations were performed before the experiments and at regular
intervals. The energies were calculated according to the method given
in Ref.~\cite{Kaufmann74}. The masses were estimated from
$M_l=A_{CN}E_l/(E_l+E_r)$, with $A_{CN}$ being the mass of the
fissioning nucleus and similarly for $M_r$.

While for the heavier targets it is rather simple to distinguish
between fission fragments and background, this is not so for silver.
For this target we therefore performed experiments with a different
setup. Again to the right of the beam was a solid state detector. To
the left we employed an ionization chamber, where the anode was
subdivided to allow for $\Delta E-E$ measurements. Because of a
Frisch grid, the signals were independent of the position of the
ionization. Position measurements were performed using a proportional
counter tube situated behind a slit in the anode. The position was
calibrated with a moveable slit between the ionization chamber and
Californium source. Energy calibration was performed as before, but
corrections had to be applied for the entrance window of the
ionization chamber.

Backgrounds were eliminated by making cuts on the scatter plots of
combinations of the following parameters: $\Delta E$, time
difference, position signals left and right of the proportional
counter tube, and the energy signal in the solid state detector and
in the ionization chamber.
\begin{figure}[!h]
\begin{center}
\includegraphics[width=0.50\textwidth]{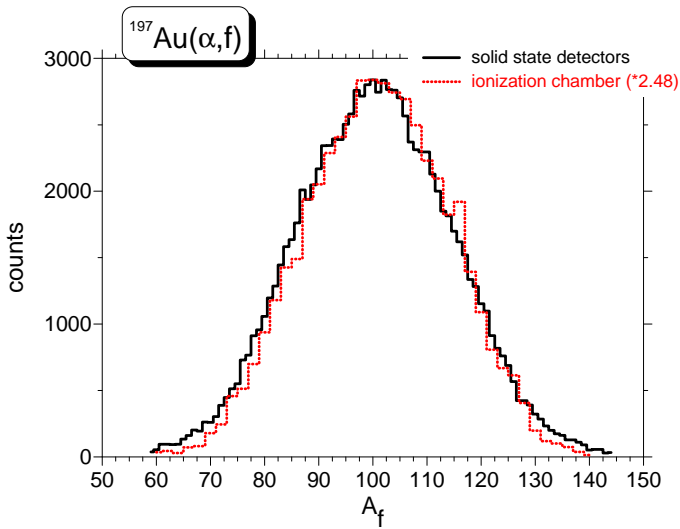}
\caption{(Color online) Comparison of the fragment distribution
measured with two solid state detectors and one solid state detector
and an ionization chamber.} \label{fig:comparison}
\end{center}
\end{figure}
In Fig.~\ref{fig:comparison} we compare the fission fragment
distributions obtained from the two methods for the case of the gold
target.

%%\section{Results}

\begin{figure}
\begin{center}
\includegraphics[width=0.5\textwidth]{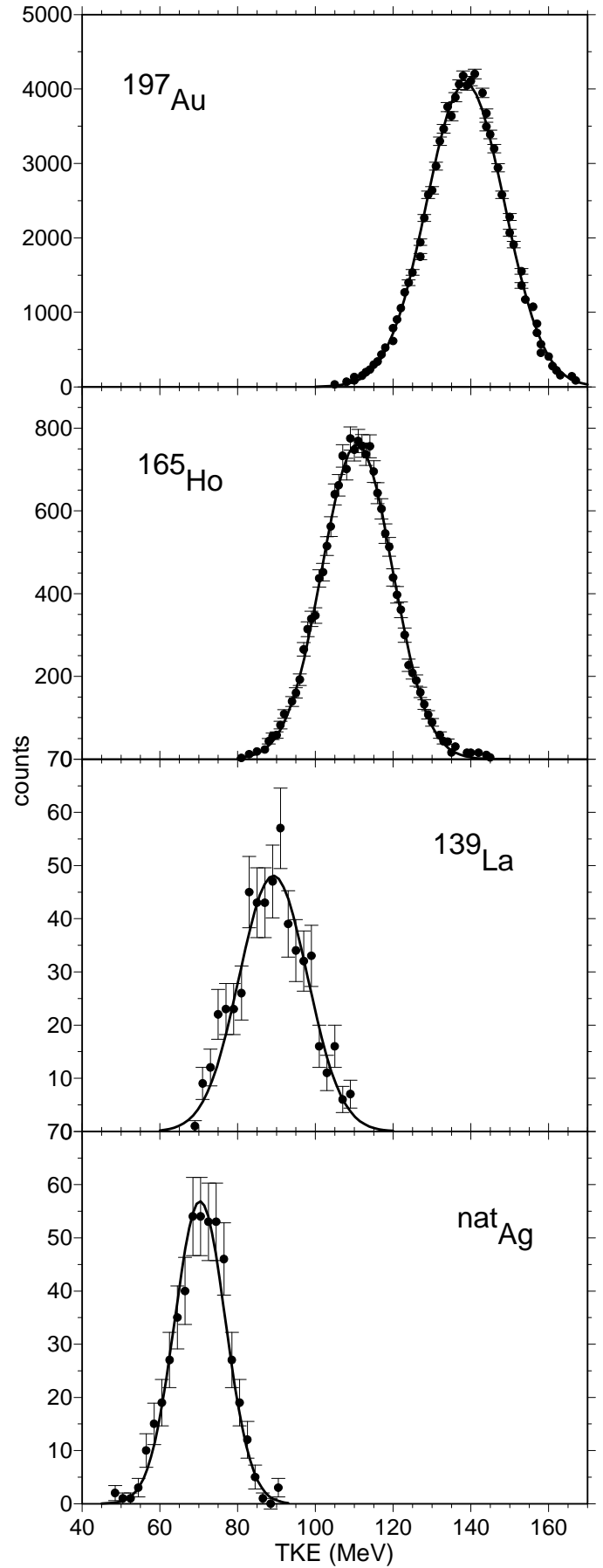}
\caption{Distributions of the total kinetic energy ($TKE$) for the
four target nuclei studied. The data are shown as dots with error
bars; fits with Gaussians are shown as solid curves. }
\label{Fig:TKE}
\end{center}
\end{figure}

\begin{figure}
\begin{center}
\includegraphics[width=0.5\textwidth]{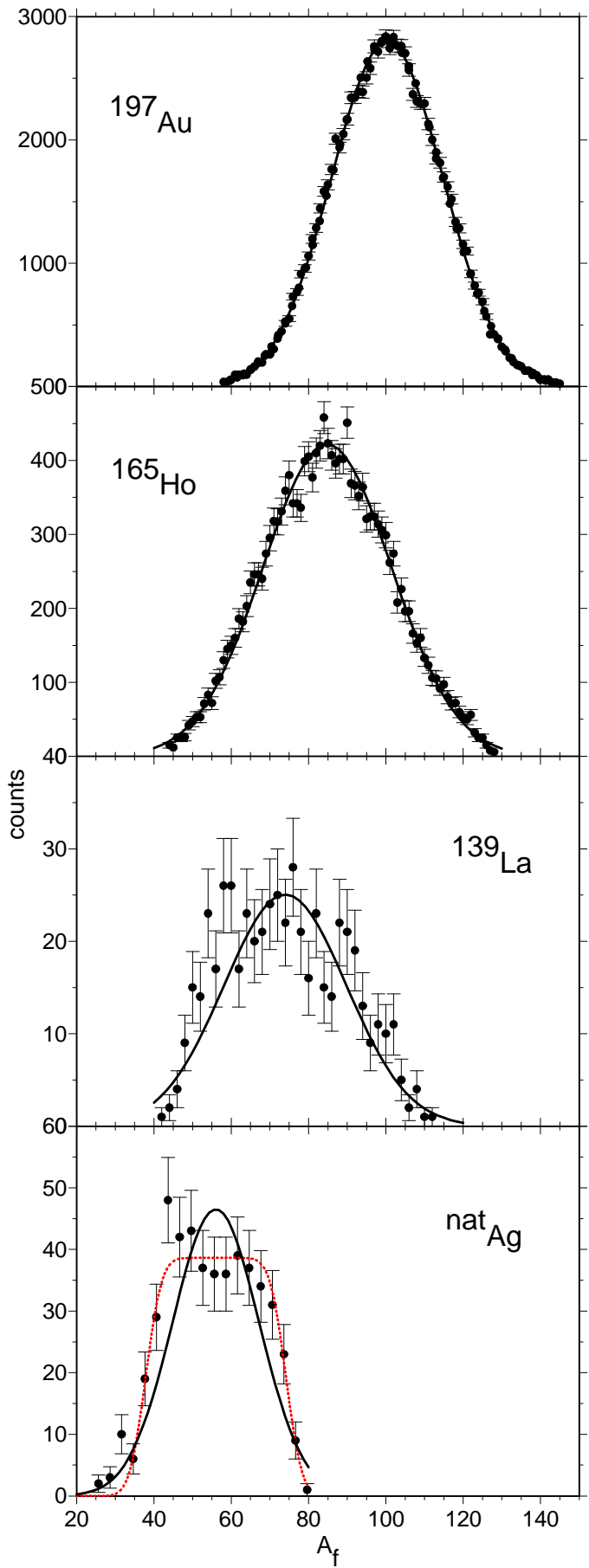}
\caption{(Color online) Fragment mass distributions for the four target nuclei
studied. The data are shown as dots with error bars; fits with
Gaussians are shown as solid curves. The fit with a symmetric double Gaussian cumulative in
the case of Ag is shown as dotted curve.} \label{fig:masses}
\end{center}
\end{figure}

The measured distributions of the total kinetic energies and the
masses of the two fragments are shown in Figs.~\ref{Fig:TKE} and
\ref{fig:masses}. In general the spectra have Gaussian shapes, as
expected for high energy fission where shell effects are unimportant.
The only exception is the mass distribution in the case of the silver
target where the distribution shows more of a box-like form. Out of
curiosity we have fitted a symmetric double Gaussian cumulative to
the data and found that it gives a much better representation. This
is also shown in Fig.~\ref{fig:masses}.


The measurements were so far analyzed as if the measurements were for
the fissioning system. These of course provide the most interesting
information. However, the experiment records events after emission of
particles before the fragments have reached the detectors. The best
choice for a necessary correction would have been through the
measurement of the corresponding particle spectra, but this was not
done. However, we performed a measurement of the correlation angle
between the two fission fragments in the case of the gold target and
found an almost complete linear momentum transfer from the projectile
to the fissioning system similar to that in Ref.~\cite{Viola74}. Furthermore we calculated the mean momentum carried away near the forward direction
by protons and neutrons during the pre-equilibrium phase by making
use of the exciton model with standard input
parameters~\cite{Machner85}. In the case of the gold target, neutrons
carry away $\approx$255~MeV/$c$ but with 0.22 neutrons per incident
$\alpha$-particle. For protons, the mean momentum is higher,
$\approx$285~MeV/$c$, but the rate is only 0.14 protons per
$\alpha$-particle.  We, therefore, assume that the pre-fission
neutrons are mainly emitted isotropically.

The average number of pre-fission neutrons $\nu_\text{pre}$ was
calculated with the ALICE code~\cite{Blann91}. Here we employed
ratios  for the level density parameters $a_f/a_n$ varying between 1.03 in the case of the gold target and 1.1 for silver. We tested this method by
studying the compound nucleus $^{208}$Po by using the results of
Cuninghame \emph{et al.}~\cite{Cuninghame80} who measured the total
number of emitted neutrons from $^{208}$Po by radiochemical methods.
They then calculated the number of pre-fission neutrons with the
ALICE code and extracted the average number of post-fission neutrons.
The latter are emitted from the excited fragments as $s$-waves and
thus do not on the average change the fragment's velocity. Their
results agree with the findings of Cheifetz \emph{et
al.}~\cite{Cheifetz70} who measured the pre-fission and the
post-fission neutrons directly at some higher energies for
$^{210}$Po. It is interesting to note that the number of post-fission
neutrons is only weakly energy dependent. The numbers for the energy and mass distributions, before and after corrections for neutron emission, are
given in Table~\ref{Tab:1}.

\begin{table}\centering
\caption{Widths and centroids of the measured distributions for the
different target nuclei. The average numbers of pre-fission neutrons $<\nu_{pre}>$ and post-fission neutrons $<\nu_{post}>$ as well as the quantities corrected for pre- and
post-fission neutron emission are also given (indicated by an
asterisk). The last row is the estimation of the mean total kinetic
energy according to the Viola method~\cite{Viola85}.}\label{Tab:1}
\begin{ruledtabular}
\begin{tabular}{c|cccc}
%%\hline
{target} & {$^{nat}$Ag} & {$^{139}$La} & {$^{165}$Ho} & {$^{197}$Au} \\
\hline
$\sigma_m$ (u) & 11.2$\pm$0.7 & 15.5$\pm$0.8 & 16.6$\pm$0.3 & 14.4$\pm$0.1 \\
$<$TKE$>$ (MeV) & 70.4$\pm$0.3 & 89.2$\pm$0.4 & 110.7$\pm$0.1 & 138.6$\pm$0.1 \\
$\sigma_{TKE}$ (MeV) & 6.5$\pm$0.2 & 8.8$\pm$0.4 & 8.2$\pm$0.1 & 10.0$\pm$0.1 \\
\hline
$<\nu_{pre}>$ & 1.0 & 1.5 & 2.7 & 7.0 \\
$<\nu_{post}>$ & 5.6 & 6.2 & 6.6 & 4.1 \\
\hline
$\sigma^*_m$ (u) & 10.2 & 15.0 &16.2  & 13.8 \\
$<$TKE$^*>$ (MeV) &71.2  & 91.9 &114.2  &138.9  \\
$\sigma^*_{TKE}$ (MeV) & 8.0 & 10.1 &9.6  &11.3
\\
\hline
$<$TKE$>_V$ & 66.6 & 86.4 & 109.7 & 140.7 \\
%%\hline
\end{tabular}
\end{ruledtabular}
\end{table}

In the following step, the influence of post-fission neutrons on the
distributions is taken into account following the method of
Ref.~\cite{Plasil66}. The resulting values are also given in
Table~\ref{Tab:1}. Here we have corrected an overestimation of the
pulse height defect in the case of the silver target, which is
approximately 1~MeV.

Finally we performed one more comparison to test the procedure. Fission is treated in ALICE in the framework of the rotating liquid drop model \cite{CPS}. We repeat the calculations for the case of the gold target with a combined cascade model code \cite{Cugnon97} and the statistical code GEM \cite{Furihata00}.  Contrary to the ALICE calculation, GEM, based on the RAL model \cite{Atchison}, treats fission on empirical parameterizations.  This limits fission to nuclei with $Z>70$. The calculation gives a mean number of emitted neutrons of 11.1 which is a nice agreement to the sum 11.6 in Table \ref{Tab:1}.

Viola and co-workers~\cite{Viola85} found a remarkable correlation
between the kinetic energy release and the Coulomb parameter
$Z^2/A^{1/3}$ (Viola-systematics).  The corresponding values are also given in the table.
The present values are slightly larger than those predicted by the systematics except for the case of gold.
\begin{table}
\centering
\caption{Cross section for fission for the different target nuclei. Also given are estimates for the fission barriers obtained by the linear dependence of the fission parameter (denoted by (I)) and on Eq. \ref{equ:barrier} (denoted by (II)). }\label{Tab:2}
\begin{ruledtabular}
\begin{tabular}{cccc}
%%\hline
target & $\sigma_{\text{fiss}}$ (mb)& $B_f$ (MeV) (I)&  $B_f$ (MeV) (II)\\
\hline
$^{\text{nat}}$Ag & 0.030$\pm$0.007 &38.8 &49.1\\
$^{139}$La & 0.007$\pm$0.001&49.5&62.8 \\
$^{165}$Ho & 0.600$\pm$0.050 &40.8&45.4\\
$^{197}$Au & 128$\pm$18 &26.9&25.7\\
%%\hline
\end{tabular}
\end{ruledtabular}
\end{table}

We convert the measured count rates into cross sections by making use
of the target thicknesses and the incident flux. The results are
given in Table~\ref{Tab:2}.
\begin{figure}
\begin{center}
\includegraphics[width=0.5\textwidth]{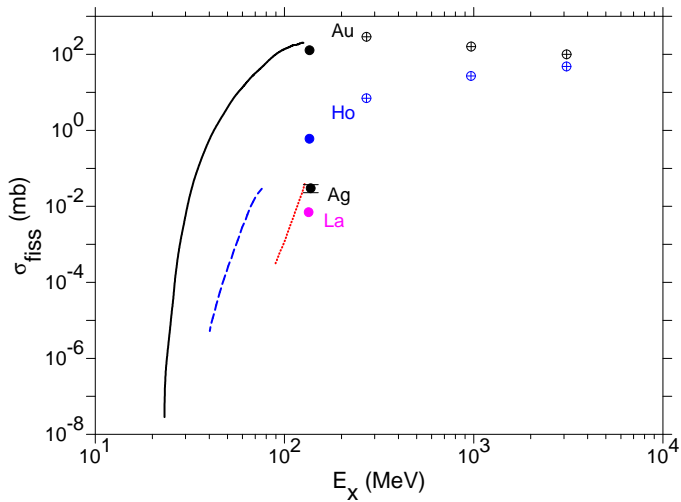}
\caption{(Color online) Excitation function for $\alpha$-particle induced fission.
The present data are shown as full dots with error bars with the
target nucleus being indicated. The curves are calculations from
Moretto~\cite{Moretto74} for In (dotted), Tm (dashed) and Au (solid).
The results from Ref.~\cite{Klotz-Engmann89} for Au and Ho are shown
by crossed circles.} \label{Fig:Exfu}
\end{center}
\end{figure}

In Fig.~\ref{Fig:Exfu} our results are compared with other data. Also
shown are calculations for nearby targets which describe remarkably
well the experimental fission cross sections at lower energies. Our
present data fill a gap between these lower energy data and results
at higher energies~\cite{Klotz-Engmann89}. It is noteworthy that the cross section in the case if the lanthanum target is the smallest.

\begin{figure}
\begin{center}
\includegraphics[width=0.5\textwidth]{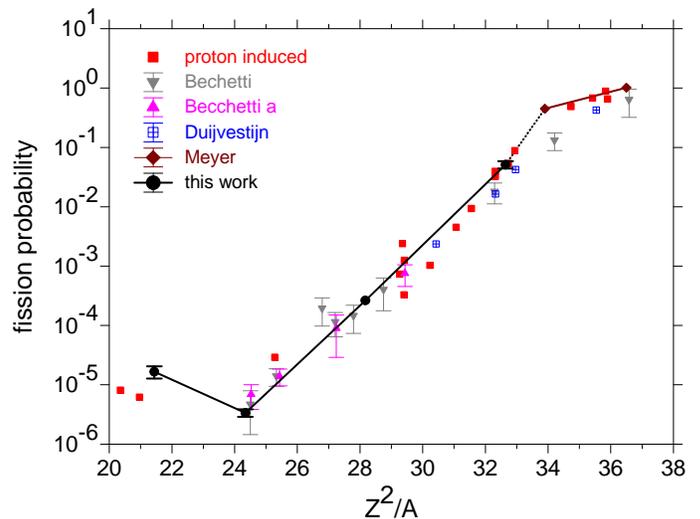}
\caption{(Color online) The fission probability as a function of the fissility
parameter. The dots with error bars are the present results and
diamonds are from Ref.~\cite{Meyer79}. The lines are to guide the
eye. The squares are for proton-induced fission at energies 150 to
200~MeV~\cite{Barashenkov72, Shigaev73}. The other data shown were measured with 190~MeV protons: triangles down \cite{Becchetti83}, triangles up \cite{Becchetti83a}, and those shown by crossed squares were measured by radiochemical
methods~\cite{Duijvestijn99}.} \label{Fig:fissility}
\end{center}
\end{figure}

Another comparison is made on the basis of the fission probability as
a function of the fissility parameter $Z^2/A$. The only data for
140~MeV $\alpha$-particles are for bismuth and
uranium~\cite{Meyer79}. The fission probability for bismuth is almost
an order of magnitude larger than that for gold and for uranium it is
about unity. Since no data exist for $\alpha$-particle induced
fission for lower masses, we compare the present data with those from
proton-induced reactions at energies close by. There is a remarkable
agreement between the results for the two entrance channels, esp. the minimum in the vicinity of $Z^2/A=24$ is visible in both reactions. The fission probability shows exponential dependencies: one slope for $24\leq Z^2/A \leq 33$ and another one for $34.5\leq Z^2/A$.

Obviously the variation of the fission probability reflects a variation of the fission barrier. In order to obtain a rough estimate we have extrapolated the Thomas-Fermi approach \cite{Myers99} to lower fissility parameters $X$, which depends not only on $Z^2/A$ but also on the symmetry energy. For a range $30\le X\le 34.14$ the original work found a linear relation of the reduced fission barrier $F(X)$ from fits to experiments. We have assumed this dependence to be valid also for smaller $X$. Alternatively we have fitted functions to all available experimental data and found $F(X)=\exp(1.6248-5.504E-05*X^3)$. The barrier is then
\begin{equation}\label{equ:barrier}
B_f(A,Z) = S(A,Z)*F(X)
\end{equation}
with $S(A,Z)$ approximately the nominal surface energy of a nucleus. The results with both methods are also given in Table \ref{Tab:2}. While the two methods give similar results in case of the gold target they diverge more to the lighter systems. However, both methods give a maximum for the Lanthanum target which corresponds to a minimum to fission probability.

%%\section{Conclusions}
In summary, we have measured the binary fission of four nuclei from
silver to gold induced by 140~MeV $\alpha$-particles. The
distributions obtained for fragment masses and total kinetic energies
were corrected for pre- and post-fission neutron emission. The mean
values of the total kinetic energies are close to those predicted by
the Viola systematics. The present data fill a gap or extend smoothly
fission yields to higher energies. The measured fission probabilities
show a distinctly different behavior from those observed for very
heavy nuclei. This is in agreement with fission studies of
proton-induced reactions at slightly higher beam energies. The
increase in the relative width of the mass distribution from
lanthanum on, as predicted by Ref.~\cite{Nix69} on the basis of the
liquid drop model, is not seen here. However, $\sigma^{*}_m/A_{CN}$
decreases from lanthanum to silver.

The minimum of the fission probability around $Z^2/A$=20 predicted
by calculations within the liquid drop model \cite{Nix66} and in a more refined model~\cite{Iljinov78,
Ivanov95} is not seen here but a minimum in the vicinity of $Z^2/A$=24. This is in agreement with proton induced fission at energies close by and in photo-fission \cite{METHASIRI71, Emma76}. A crude estimate of fission barriers in a Thomas-Fermi approach yields a maximum for that value.   The fragment
mass distribution in the case of the silver target shows an almost
rectangular shape while the other mass distributions look Gaussian.
Such a behavior was found in fission of $^{232}$Th with 190~MeV
protons~\cite{Duijvestijn99}. A possible explanation is that after
neutron emission the low excited system still undergoes fission which
is then asymmetric. This was also found in antiproton annihilation on
$^{238}$U~\cite{Machner92}. However, such an explanation is very
unlikely for silver-like compound nuclei. Therefore this case needs
further investigation.

%%%%%%%%%%%%%%%%%%%%%%%%%%%%%%%%%%%%%%%%%%%%%%%%
% Begin References
%%%%%%%%%%%%%%%%%%%%%%%%%%%%%%%%%%%%%%%%%%%%%%%%

%%%%%%%%%%%%%%%%%%%%%%%%%%%%%%%%%%%%%%%%%%%%%%%%%
% End References
%%%%%%%%%%%%%%%%%%%%%%%%%%%%%%%%%%%%%%%%%%%%%%%%%

%%%%%%%%%%%%%%%%%%%%%%%%%%%%%%%%%%%%%%%%%%%%%%%%
% End contact information
%%%%%%%%%%%%%%%%%%%%%%%%%%%%%%%%%%%%%%%%%%%%%%%%


\begin{thebibliography}{22}
\expandafter\ifx\csname natexlab\endcsname\relax\def\natexlab#1{#1}\fi
\expandafter\ifx\csname bibnamefont\endcsname\relax
  \def\bibnamefont#1{#1}\fi
\expandafter\ifx\csname bibfnamefont\endcsname\relax
  \def\bibfnamefont#1{#1}\fi
\expandafter\ifx\csname citenamefont\endcsname\relax
  \def\citenamefont#1{#1}\fi
\expandafter\ifx\csname url\endcsname\relax
  \def\url#1{\texttt{#1}}\fi
\expandafter\ifx\csname urlprefix\endcsname\relax\def\urlprefix{URL }\fi
\providecommand{\bibinfo}[2]{#2}
\providecommand{\eprint}[2][]{\url{#2}}

\bibitem[{\citenamefont{Vandenboosch and Huizenga}(1973)}]{Vanden73}
\bibinfo{author}{\bibfnamefont{R.}~\bibnamefont{Vandenboosch}}
  \bibnamefont{and} \bibinfo{author}{\bibfnamefont{J.~R.}
  \bibnamefont{Huizenga}}, \emph{\bibinfo{title}{Nuclear Fission}}
  (\bibinfo{publisher}{Academic Press}, \bibinfo{address}{New York and London},
  \bibinfo{year}{1973}).

\bibitem[{\citenamefont{Wagemans}(1991)}]{Wagemans91}
\bibinfo{author}{\bibfnamefont{C.}~\bibnamefont{Wagemans}},
  \emph{\bibinfo{title}{The Nuclear Fission Process}} (\bibinfo{publisher}{CRC
  Press}, \bibinfo{address}{Boca Raton}, \bibinfo{year}{1991}).

\bibitem[{\citenamefont{von Oertzen et~al.}(2008)\citenamefont{von Oertzen,
  Zherebchevsky, Gebauer, Schulz, Thummerer, Kamanin, Royer, and
  Wilpert1}}]{Oertzen08}
\bibinfo{author}{\bibfnamefont{W.}~\bibnamefont{von Oertzen}}
%  \bibinfo{author}{\bibfnamefont{V.}~\bibnamefont{Zherebchevsky}},
%  \bibinfo{author}{\bibfnamefont{B.}~\bibnamefont{Gebauer}},
%  \bibinfo{author}{\bibfnamefont{C.}~\bibnamefont{Schulz}},
%  \bibinfo{author}{\bibfnamefont{S.}~\bibnamefont{Thummerer}},
%  \bibinfo{author}{\bibfnamefont{D.}~\bibnamefont{Kamanin}},
% \bibinfo{author}{\bibfnamefont{G.}~\bibnamefont{Royer}}, \bibnamefont{and}
% \bibinfo{author}{\bibfnamefont{T.}~\bibnamefont{Wilpert1}},
  \bibnamefont{et~al.},
  \bibinfo{journal}{Phys. Rev.} \textbf{\bibinfo{volume}{C 78}},
  \bibinfo{pages}{044615} (\bibinfo{year}{2008}).

\bibitem[{\citenamefont{Businaro and Gallone}(1957)}]{Businaro57}
\bibinfo{author}{\bibfnamefont{U.~L.} \bibnamefont{Businaro}} \bibnamefont{and}
  \bibinfo{author}{\bibfnamefont{S.}~\bibnamefont{Gallone}},
  \bibinfo{journal}{Nuovo Cimento} \textbf{\bibinfo{volume}{5}},
  \bibinfo{pages}{315} (\bibinfo{year}{1957}).

\bibitem[{\citenamefont{Nix and Sassi}(1966)}]{Nix66}
\bibinfo{author}{\bibfnamefont{J.~R.} \bibnamefont{Nix}} \bibnamefont{and}
  \bibinfo{author}{\bibfnamefont{E.}~\bibnamefont{Sassi}},
  \bibinfo{journal}{Nucl. Phys.} \textbf{\bibinfo{volume}{81}},
  \bibinfo{pages}{61} (\bibinfo{year}{1966}).

\bibitem[{\citenamefont{Schmitt et~al.}(1965)\citenamefont{Schmitt, Kiker, and
  Williams}}]{Schmitt65}
\bibinfo{author}{\bibfnamefont{H.~W.} \bibnamefont{Schmitt}},
  \bibinfo{author}{\bibfnamefont{W.~E.} \bibnamefont{Kiker}}, \bibnamefont{and}
  \bibinfo{author}{\bibfnamefont{C.~W.} \bibnamefont{Williams}},
  \bibinfo{journal}{Phys. Rev.} \textbf{\bibinfo{volume}{137}},
  \bibinfo{pages}{37} (\bibinfo{year}{1965}).

\bibitem[{\citenamefont{Kaufmann et~al.}(1974)}]{Kaufmann74}
\bibinfo{author}{\bibfnamefont{B.}~\bibnamefont{Kaufmann}}
  \bibinfo{author}{\bibnamefont{et~al.}}, \bibinfo{journal}{Nucl. Inst. \&
  Meth.} \textbf{\bibinfo{volume}{115}}, \bibinfo{pages}{47}
  (\bibinfo{year}{1974}).

\bibitem{Viola74}
V. E. Viola, C. T. Roche, W. G. Meyer, R. G. Clark, Phys.\ Rev.\ C \textbf{10}, 2416 (1974).


\bibitem[{\citenamefont{Machner}(1985)}]{Machner85}
\bibinfo{author}{\bibfnamefont{H.}~\bibnamefont{Machner}},
  \bibinfo{journal}{Phys. Rep.} \textbf{\bibinfo{volume}{127}},
  \bibinfo{pages}{309} (\bibinfo{year}{1985}).

\bibitem[{\citenamefont{Blann}(1991)}]{Blann91}
\bibinfo{author}{\bibfnamefont{M.}~\bibnamefont{Blann}}, \bibinfo{type}{Tech.
  Rep.} \bibinfo{number}{LLNL Report No. UCRL-JC-},
  \bibinfo{institution}{Lawrence Livermore National Laboratory}
  (\bibinfo{year}{1991}).

\bibitem[{\citenamefont{Cuninhame et~al.}(1980)}]{Cuninghame80}
\bibinfo{author}{\bibfnamefont{J.~G.} \bibnamefont{Cuninhame}}
\bibinfo{author}{\bibnamefont{et~al.}}, in
  \emph{\bibinfo{booktitle}{Physics and Chemistry of Fission 1979}}
  (\bibinfo{publisher}{International Atomic Energy Agency},
  \bibinfo{address}{Vienna}, \bibinfo{year}{1980}), vol.~\bibinfo{volume}{I},
  p. \bibinfo{pages}{551}.

\bibitem[{\citenamefont{Cheifetz et~al.}(1970)\citenamefont{Cheifetz, Fraenkel,
  Galin, Lefort, P\'{e}ter, and Tarrago}}]{Cheifetz70}
\bibinfo{author}{\bibfnamefont{E.}~\bibnamefont{Cheifetz}},
  \bibinfo{author}{\bibfnamefont{Z.}~\bibnamefont{Fraenkel}},
  \bibinfo{author}{\bibfnamefont{J.}~\bibnamefont{Galin}},
  \bibinfo{author}{\bibfnamefont{M.}~\bibnamefont{Lefort}},
  \bibinfo{author}{\bibfnamefont{J.}~\bibnamefont{P\'{e}ter}},
  \bibnamefont{and} \bibinfo{author}{\bibfnamefont{X.}~\bibnamefont{Tarrago}},
  \bibinfo{journal}{Phys. Rev.} \textbf{\bibinfo{volume}{C 2}},
  \bibinfo{pages}{256} (\bibinfo{year}{1970}).

\bibitem[{\citenamefont{Plasil et~al.}(1966)}]{Plasil66}
\bibinfo{author}{\bibfnamefont{F.}~\bibnamefont{Plasil}}
  \bibinfo{author}{\bibnamefont{et~al.}}, \bibinfo{journal}{Phys. Rev.}
  \textbf{\bibinfo{volume}{142}}, \bibinfo{pages}{696} (\bibinfo{year}{1966}).

\bibitem{CPS} S. Cohen, F. Plasil, W. J. Swiatecki, Ann. Phys. \textbf{82}, 557 (1974).

\bibitem{Cugnon97} J. Cugnon, C. Volant, and S. Vuillier, Nucl. Phys. \textbf{A620}, 475 (1997).

\bibitem{Furihata00} S. Furihata, Nucl. Instrum. Methods Phys. Res. \textbf{B 171}, 251 (2000).

\bibitem{Atchison} F. Atchison, "A Treatment of Fission for HETC," in Intermediate Energy Nuclear
Data: Models and Codes, pp. 199–218, Proc. of a Specialists's Meeting, May 30–June 1, 1994, Issy-Les-Moulineaux, France, OECD, Paris, France (1994).
\bibitem[{\citenamefont{Viola et~al.}(1985)\citenamefont{Viola, Kwiatkowski,
  and Walker}}]{Viola85}
\bibinfo{author}{\bibfnamefont{V.~E.} \bibnamefont{Viola}},
  \bibinfo{author}{\bibfnamefont{K.}~\bibnamefont{Kwiatkowski}},
  \bibnamefont{and} \bibinfo{author}{\bibfnamefont{M.}~\bibnamefont{Walker}},
  \bibinfo{journal}{Phys. Rev.} \textbf{\bibinfo{volume}{C 31}},
  \bibinfo{pages}{1550} (\bibinfo{year}{1985}).

\bibitem[{\citenamefont{Moretto}(1974)}]{Moretto74}
\bibinfo{author}{\bibfnamefont{L.~G.} \bibnamefont{Moretto}}, in
  \emph{\bibinfo{booktitle}{Physics and Chemistry of Fission 1973}}
  (\bibinfo{publisher}{International Atomic Energy Agency},
  \bibinfo{address}{Vienna}, \bibinfo{year}{1974}), vol.~\bibinfo{volume}{I},
  p. \bibinfo{pages}{329}.

\bibitem[{\citenamefont{Klotz-Engmann et~al.}(1989)}]{Klotz-Engmann89}
\bibinfo{author}{\bibfnamefont{G.}~\bibnamefont{Klotz-Engmann}}
\bibinfo{author}{\bibnamefont{et~al.}},
  \bibinfo{journal}{Nucl. Phys.} \textbf{\bibinfo{volume}{A 499}},
  \bibinfo{pages}{392} (\bibinfo{year}{1989}).

\bibitem[{\citenamefont{Meyer et~al.}(1979)\citenamefont{Meyer, Viola, Clark,
  Read, and Theus}}]{Meyer79}
\bibinfo{author}{\bibfnamefont{W.~G.} \bibnamefont{Meyer}}
%  \bibinfo{author}{\bibfnamefont{V.~E.} \bibnamefont{Viola}},
% \bibinfo{author}{\bibfnamefont{R.~G.} \bibnamefont{Clark}},
% \bibinfo{author}{\bibfnamefont{S.~M.} \bibnamefont{Read}}, \bibnamefont{and}
% \bibinfo{author}{\bibfnamefont{R.~B.} \bibnamefont{Theus}},
  \bibnamefont{et~al.},
  \bibinfo{journal}{Phys. Rev.} \textbf{\bibinfo{volume}{C 20}},
  \bibinfo{pages}{1716} (\bibinfo{year}{1979}).

\bibitem[{\citenamefont{Barashenkov and Toneev}(1972)}]{Barashenkov72}
\bibinfo{author}{\bibfnamefont{V.~S.} \bibnamefont{Barashenkov}}
  \bibnamefont{and} \bibinfo{author}{\bibfnamefont{V.~D.}
  \bibnamefont{Toneev}}, \emph{\bibinfo{title}{Interaction of high energy
  particles and nuclei with nuclei}} (\bibinfo{publisher}{Atomizdat Moscow},
  \bibinfo{year}{1972}), p. \bibinfo{pages}{566}.

\bibitem[{\citenamefont{Shigaev et~al.}(1973)\citenamefont{Shigaev, Bychkov,
  Lomanov, Obukhov, Perfilov, Shilichuk, and Yakolev}}]{Shigaev73}
\bibinfo{author}{\bibfnamefont{O.~E.} \bibnamefont{Shigaev}}
% \bibinfo{author}{\bibfnamefont{V.~S.} \bibnamefont{Bychkov}},
% \bibinfo{author}{\bibfnamefont{M.~F.} \bibnamefont{Lomanov}},
% \bibinfo{author}{\bibfnamefont{A.~I.} \bibnamefont{Obukhov}},
% \bibinfo{author}{\bibfnamefont{N.~A.} \bibnamefont{Perfilov}},
% \bibinfo{author}{\bibfnamefont{G.~G.} \bibnamefont{Shilichuk}},
% \bibnamefont{and} \bibinfo{author}{\bibfnamefont{R.~M.}
% \bibnamefont{Yakolev}},
   \bibnamefont{et~al.},
\bibinfo{journal}{Radium Institut Leningrad report}
  \textbf{\bibinfo{volume}{RI-17}} (\bibinfo{year}{1973}).

\bibitem{Becchetti83}
F. D. Becchetti, J. J\"{a}necke, P. Lister, K. Kwiatowski, H. Karwowski, S. Zhou, Phys. Rev. \textbf{C 28}, 276 (1983).

\bibitem{Becchetti83a}
F. D. Becchetti, P. Lister, J. J\"{a}necke, A. Nadasen, K. Kwiatowski, H. Karwowski, K. Hicks, Bull. Am. Phys. Soc \textbf{28}, 698 (1983).

\bibitem[{\citenamefont{Duijvestijn et~al.}(1999)\citenamefont{Duijvestijn,
  Koening, Beijers, Ferrari, Gastal, van Klinken, and
  Ostendorf}}]{Duijvestijn99}
\bibinfo{author}{\bibfnamefont{M.~C.} \bibnamefont{Duijvestijn}}
%  \bibinfo{author}{\bibfnamefont{A.~J.} \bibnamefont{Koening}},
%  \bibinfo{author}{\bibfnamefont{J.~P.~M.} \bibnamefont{Beijers}},
%  \bibinfo{author}{\bibfnamefont{A.}~\bibnamefont{Ferrari}},
%  \bibinfo{author}{\bibfnamefont{M.}~\bibnamefont{Gastal}},
%  \bibinfo{author}{\bibfnamefont{J.}~\bibnamefont{van Klinken}},
%  \bibnamefont{and} \bibinfo{author}{\bibfnamefont{R.~W.}
%  \bibnamefont{Ostendorf}},
   \bibnamefont{et~al.},
  \bibinfo{journal}{Phys. Rev.}
  \textbf{\bibinfo{volume}{C 59}}, \bibinfo{pages}{776} (\bibinfo{year}{1999}).

\bibitem{Myers99}
W. D. Myers and W. J. Swi\c{a}tecki, Phys. Rev. \textbf{C 60}, 014606 (1999).

\bibitem{Nix69}
J. R. Nix, Nucl. Phys. \textbf{A 130}, 241 (1969).
\bibitem[{\citenamefont{Iljinov et~al.}(1978)\citenamefont{Iljinov, Cherepanov,
  and Chigrinov}}]{Iljinov78}
\bibinfo{author}{\bibfnamefont{A.~S.} \bibnamefont{Iljinov}},
  \bibinfo{author}{\bibfnamefont{E.~A.} \bibnamefont{Cherepanov}},
  \bibnamefont{and} \bibinfo{author}{\bibfnamefont{S.~E.}
  \bibnamefont{Chigrinov}}, \bibinfo{journal}{Z. Physik}
  \textbf{\bibinfo{volume}{A 287}}, \bibinfo{pages}{37} (\bibinfo{year}{1978}).

\bibitem[{\citenamefont{Ivanov et~al.}(1995)}]{Ivanov95}
\bibinfo{author}{\bibfnamefont{D.~I.} \bibnamefont{Ivanov}}
  \bibinfo{author}{\bibnamefont{et~al.}}, \bibinfo{journal}{Z. Physik}
  \textbf{\bibinfo{volume}{A 352}}, \bibinfo{pages}{191}
  (\bibinfo{year}{1995}).

\bibitem{METHASIRI71}
T. Methasiri and S. A. E. Johansson, Nucl. Phys. \textbf{A 167}, 97 (1971).

\bibitem{Emma76}
V. Emma, S. Lo Nigro and C. Milone, Nucl. Phys. \textbf{A 257}, 438 (1976).

\bibitem[{\citenamefont{Machner et~al.}(1992)\citenamefont{Machner, Jun, and.
  Protic, Daniel, von Egidy, Hartmann, Kanert, Markiel, Plendl, Ziock
  et~al.}}]{Machner92}
\bibinfo{author}{\bibfnamefont{H.}~\bibnamefont{Machner}}
%  \bibinfo{author}{\bibfnamefont{S.}~\bibnamefont{Jun}},
%  \bibinfo{author}{\bibfnamefont{G.~R.} \bibnamefont{and. Protic}},
%  \bibinfo{author}{\bibfnamefont{H.}~\bibnamefont{Daniel}},
%  \bibinfo{author}{\bibfnamefont{T.}~\bibnamefont{von Egidy}},
%  \bibinfo{author}{\bibfnamefont{F.~J.} \bibnamefont{Hartmann}},
%  \bibinfo{author}{\bibfnamefont{W.}~\bibnamefont{Kanert}},
%  \bibinfo{author}{\bibfnamefont{W.}~\bibnamefont{Markiel}},
%  \bibinfo{author}{\bibfnamefont{H.~S.} \bibnamefont{Plendl}},
% \bibinfo{author}{\bibfnamefont{K.}~\bibnamefont{Ziock}},
  \bibnamefont{et~al.}, \bibinfo{journal}{Z. Physik} \textbf{\bibinfo{volume}{A
  343}}, \bibinfo{pages}{73} (\bibinfo{year}{1992}).

\end{thebibliography}
\end{document}